\documentclass[aps,prl,twocolumn,superscriptaddress,showpacs]{revtex4-1}
\usepackage{graphicx}
\usepackage{siunitx}
\usepackage{amssymb}
\begin{document}

\title{Structural Order and Melting of a Quasi-One-Dimensional Electron System}


\author{David G. Rees}
\email[]{drees@nctu.edu.tw}
\affiliation{NCTU-RIKEN Joint Research Laboratory, Institute of Physics, National Chiao Tung University, Hsinchu 300, Taiwan}
\affiliation{RIKEN CEMS, Wako 351-0198, Japan}

\author{Niyaz R. Beysengulov}
\affiliation{RIKEN CEMS, Wako 351-0198, Japan}
\affiliation{KFU-RIKEN Joint Research Laboratory, Institute of Physics, Kazan Federal University, Kazan, 420008 Russia}

\author{Yoshiaki Teranishi}
\affiliation{NCTU-RIKEN Joint Research Laboratory, Institute of Physics, National Chiao Tung University, Hsinchu 300, Taiwan}
\affiliation{Physics Division, National Center for Theoretical Sciences, Hsinchu 300,
Taiwan} 

\author{Chun-Shuo Tsao}
\affiliation{NCTU-RIKEN Joint Research Laboratory, Institute of Physics, National Chiao Tung University, Hsinchu 300, Taiwan}
\affiliation{Physics Division, National Center for Theoretical Sciences, Hsinchu 300,
Taiwan} 

\author{Sheng-Shiuan Yeh}
\affiliation{NCTU-RIKEN Joint Research Laboratory, Institute of Physics, National Chiao Tung University, Hsinchu 300, Taiwan}

\author{Shao-Pin Chiu}
\affiliation{NCTU-RIKEN Joint Research Laboratory, Institute of Physics, National Chiao Tung University, Hsinchu 300, Taiwan}

\author{Yong-Han Lin}
\affiliation{NCTU-RIKEN Joint Research Laboratory, Institute of Physics, National Chiao Tung University, Hsinchu 300, Taiwan}

\author{Dmitrii A. Tayurskii}
\affiliation{RIKEN CEMS, Wako 351-0198, Japan}
\affiliation{KFU-RIKEN Joint Research Laboratory, Institute of Physics, Kazan Federal University, Kazan, 420008 Russia}

\author{Juhn-Jong Lin}
\affiliation{NCTU-RIKEN Joint Research Laboratory, Institute of Physics, National Chiao Tung University, Hsinchu 300, Taiwan}
\affiliation{RIKEN CEMS, Wako 351-0198, Japan}
\affiliation{Department of Electrophysics, National Chiao Tung University, Hsinchu 300, Taiwan}

\author{Kimitoshi Kono}
\affiliation{NCTU-RIKEN Joint Research Laboratory, Institute of Physics, National Chiao Tung University, Hsinchu 300, Taiwan}
\affiliation{RIKEN CEMS, Wako 351-0198, Japan}
\affiliation{KFU-RIKEN Joint Research Laboratory, Institute of Physics, Kazan Federal University, Kazan, 420008 Russia}

\date{\today}
\begin{abstract}
We investigate the influence of confinement on the positional order of a quasi-1D electron system trapped on the surface of liquid helium. We find evidence that the melting of the Wigner solid (WS) depends on the confinement strength, as well as electron density and temperature. A reentrant solid-liquid-solid transition is observed for increasing electron density under constant electrostatic confinement. As the electron row number $N_y$ changes, varying commensurability results in a modulation of the WS order, even when $N_y$ is large (several tens). This is confirmed by Monte Carlo simulations.   
\end{abstract}

\maketitle
\section{I. Introduction}
For interacting particles in (quasi-)1D channels, the competing influences of temperature, interaction energy and confinement give rise to many complex phenomena. For Fermi degenerate electrons (or holes), 1D confinement results in Luttinger liquid behaviour\cite{deshpande2010electron}, Wigner crystallisation\cite{deshpande2008one} and anomalous transport close to the first quantised conductance plateau\cite{PhysRevLett.77.135}. As the confinement weakens, `zig-zag' transitions from 1 to 2 electron rows can occur\cite{Hewetal,chaplik1980instability}. Similar structural transitions can be observed directly in trapped ion experiments\cite{birkl1992multiple}. For a growing number of particle chains the quasi-1D order depends critically on commensurability, as demonstrated in experiments with colloids\cite{LeidererLayering}, dusty plasmas\cite{PhysRevLett.90.245004}, vortices in superconducting films\cite{VorticesInChannels} and electrons on liquid helium substrates\cite{PhysRevLett.109.236802, *rees2013reentrant}, as well as numerical simulations\cite{Peeters1DCrystal,PackingAndMelting,PhysRevB.84.024117}. However, the parameter range explored in such experiments is typically quite narrow, whilst simulations are limited by processor speed. Here, we gain new insights into the ordering and melting of a quasi-1D system of electrons on He by using a multigated microchannel device to tune the particle density and confinement over a wide range. In particular, we find that the strength of the lateral confinement, rather than simply the reduced width, can play a key role in determining the melting behaviour of the quasi-1D electron lattice. 

Surface-state electron (SSE) systems on liquid He substrates are ideally suited to the study of strongly correlated electron behaviour\cite{Andrei}. The typical surface density ($n_s\approx10^{13}$ m$^{-2}$) is low, ensuring that electron-electron interactions are purely Coulombic. In 2D, the electrons form a triangular lattice, the Wigner solid (WS), with increasing $n_s$ or decreasing temperature $T$\cite{GrimesAdamsWignerCrystal}. Once the electrons become localised, the electrostatic pressure from each electron on the He beneath results in the formation of surface `dimples' that increase the system effective mass\cite{MonarkhaShikinDimple}. For the moving electron lattice, resonant Bragg-Cherenkov (BC) scattering with surface excitations (ripplons) deepens the dimple lattice and the resistive force exerted on the electron system increases\cite{DykmanRubo}. The electron velocity is then limited to that of the phase velocity of ripplons whose wavevector is commensurate with one of the reciprocal lattice vectors of the electron lattice (usually the shortest). Hence the Wigner solid transport is strongly nonlinear. When the driving force reaches a critical value, the electrons decouple from the dimple lattice and `slide' along the He surface with high velocity\cite{PhysRevLett.74.781}. This transport anomaly is a sensitive alert to the Wigner solid formation\cite{DahmMobillity}. The onset of nonlinearity in the electric conductivity is a more reliable experimental criterion for determining the Wigner solid formation than the conductivity change, which can appear smeared\cite{rees2013reentrant}.

Microchannels filled with liquid He can provide quasi-1D confinement for SSE systems\cite{PhysRevLett.87.176802,ReesPRL}. Using such devices, the melting of the Wigner solid was found to be suppressed for small electron row number $N_y$ whilst, separately, reentrant ordering of the quasi-1D lattice was observed with increasing $N_y$\cite{PhysRevB.82.201104, *Ikegami2015, PhysRevLett.109.236802, *rees2013reentrant}. Here we map structural and phase diagrams for a quasi-1D SSE system, for $1\leq N_y\lesssim 30$. The phase boundary is determined by a unique method, namely measuring the third harmonic component of the SSE current to find the onset of nonlinear transport.  We find evidence of a scaled relationship between Coulomb energy, confinement strength and temperature at the Wigner solid melting point, and observe a novel solid-liquid-solid melting behaviour as $n_s$ increases under constant electrostatic confinement. Our experiment demonstrates a uniquely sensitive electrostatic control over the positional order of a quasi-1D electron system, and so is an important step towards utilising SSE in quantum information and quantum optics applications\cite{PlatzmanandDykmanscience,LyonQubits,SchusterProposal}.

\begin{figure} 
\includegraphics[angle=0,width=0.235\textwidth]{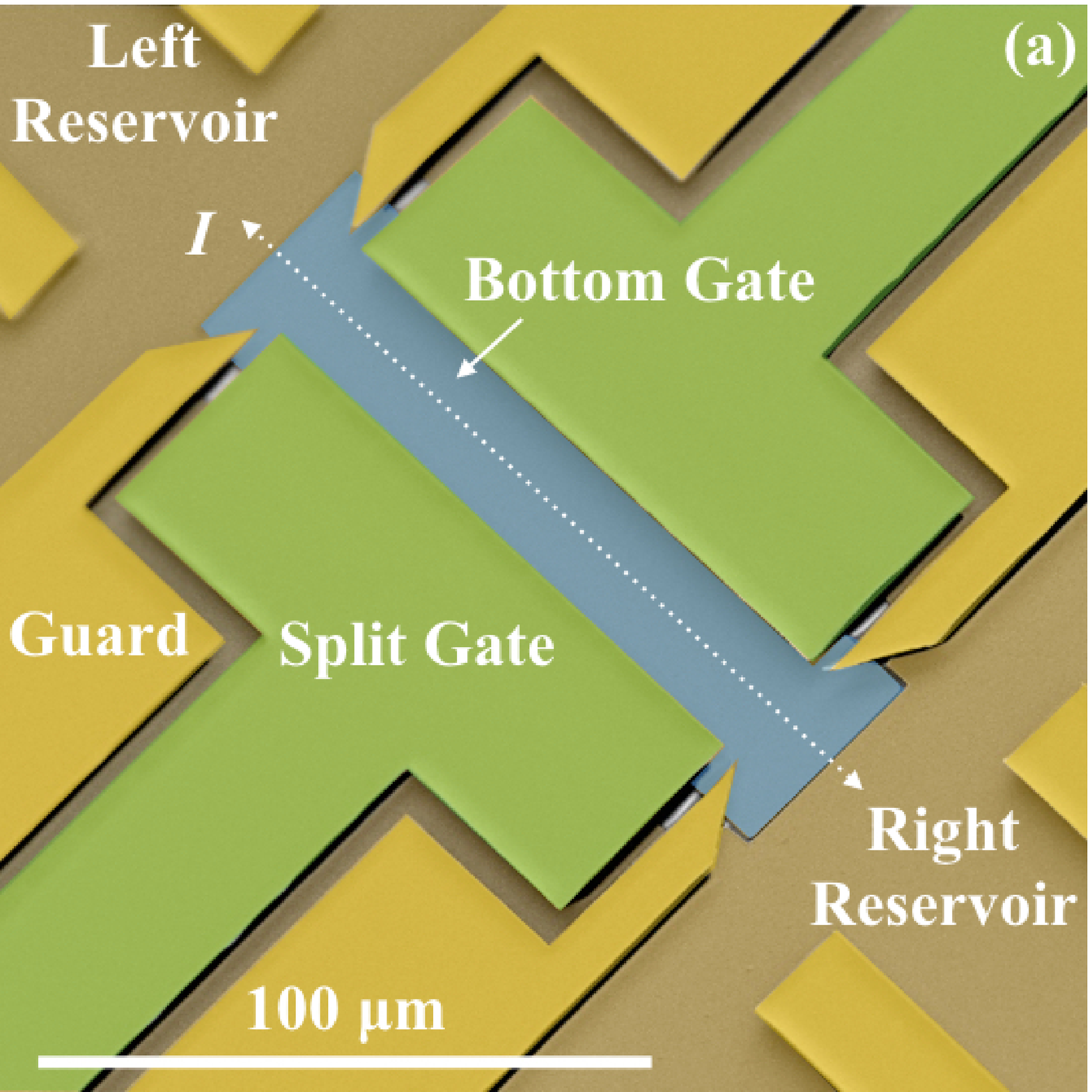}
\includegraphics[angle=0,width=0.235\textwidth]{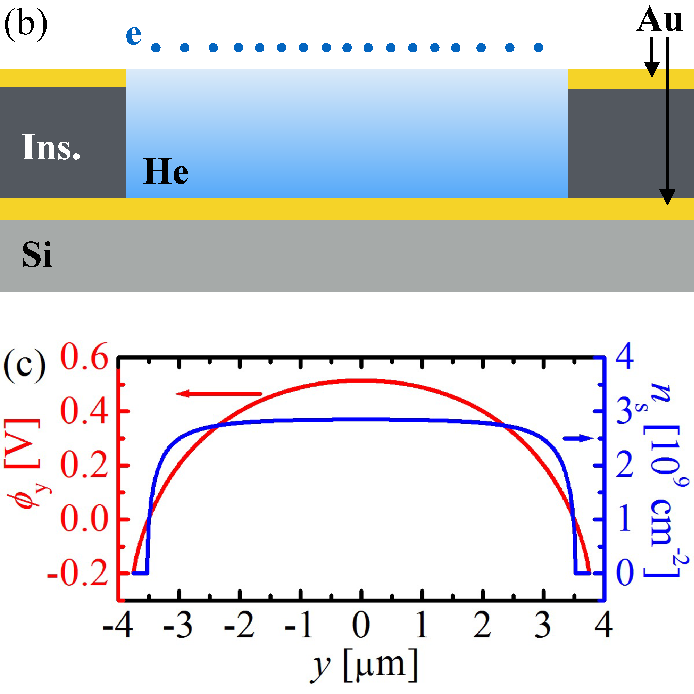}
\caption{(a) False-color scanning electron micrograph of the central microchannel in the device. (b) Schematic diagram of the microchannel cross-section. Gold layers, separated by an insulating layer, are fabricated on a silicon wafer section. (c) $\phi_y$ and $n_s$ in the central microchannel calculated by FEM for $V_{sg}=-0.2$ V and $V_{bg}=+1$ V. \label{Fig:2}}
\end{figure} 
\section{II. Results and Discussion}
\subsection{A. Sample and Method}
The device, shown in Fig. 1(a), has been described in detail elsewhere\cite{Beysengulov}. Two metal layers are separated by an insulating layer approximately 2 $\SI{}{\micro\meter}$ thick formed by hard-baked photoresist. The Guard electrode (upper layer) and Reservoir electrodes (lower layer) define two large arrays of microchannels that act as electron reservoirs. The reservoirs are connected by a smaller central microchannel, 100 $\SI{}{\micro\meter}$ long and 7.5 $\SI{}{\micro\meter}$ wide, that is formed by the Split Gate electrode (upper layer) and Bottom Gate electrode (lower layer).

Dc voltages $V_{gu}=-0.2$ V, $V_{res}=0$ V, $V_{bg}$ and $V_{sg}$ were applied to the Guard, Reservoir, Bottom Gate and Split Gate electrodes, respectively. Transport measurements were made by applying an ac voltage $V_{in}$, of frequency $f=20.2$ kHz, to the Left Reservoir electrode and measuring the ac current $I$ induced in the Right Reservoir electrode. (Note that all ac parameters are given in peak-to-peak units.) The circuit was well-described by a lumped-element RC model\cite{Iye}, which was used to extract the SSE resistance $R$. The area of the reservoirs greatly exceeds that of the central microchannel. Therefore, the number of electrons in the reservoirs, and so the electrostatic potential of the electron system $\phi_e$, can be assumed to remain constant whilst the lateral confinement potential $\phi_y$ in the central microchannel is controlled by changing $V_{bg}$ and $V_{sg}$ (Fig. 1(c)). The electron density in the reservoirs was kept low to prevent Wigner solid formation, a crucial advantage over previous experiments\cite{PhysRevLett.109.236802, *rees2013reentrant}. Finite element modelling (FEM) was used to calculate $\phi_y$, the average $n_s$ and the effective width of the electron system $w_e$ in the central microchannel for all values of $V_{bg}$ and $V_{sg}$, using $\phi_e$ and the channel depth $h$ as fitting parameters\cite{Beysengulov,hecht2012new}. $N_y$ was then estimated as $N_y=w_e\sqrt{n_s}$. Values of the angular frequency $\omega$, which describes strength of the lateral confinement, were found by fitting the parabolic function $\phi_y e=\frac{1}{2}m_e\omega^2y^2$ to the FEM results in the central region of the central microchannel, where $m_e$ is the bare electron mass.

\subsection{B. Experimental Results}
The magnitude of $I$ for varying $V_{bg}$ and $V_{sg}$ is shown in Fig. 2(a), for $V_{in}=3$ mV and $T=0.6$ K. The threshold for conductance through the central microchannel depends on both electrode voltages. As in other similar devices\cite{ReesPRL,Beysengulov}, electrons can enter the central microchannel when the maximum of the potential at its centre (or minimum, for electrons) $\phi_y^{max}=\alpha V_{bg}+\beta V_{sg}$ exceeds $\phi_e$. Examination of the conductance threshold in Fig. 2(a) yields the values $\alpha=0.60$ and $\beta=0.40$. These values are reproduced by the FEM for $h=2.20$ $\SI{}{\micro\meter}$. 

 \begin{figure} 
\begin{centering}
 \includegraphics[angle=0,width=0.45\textwidth]{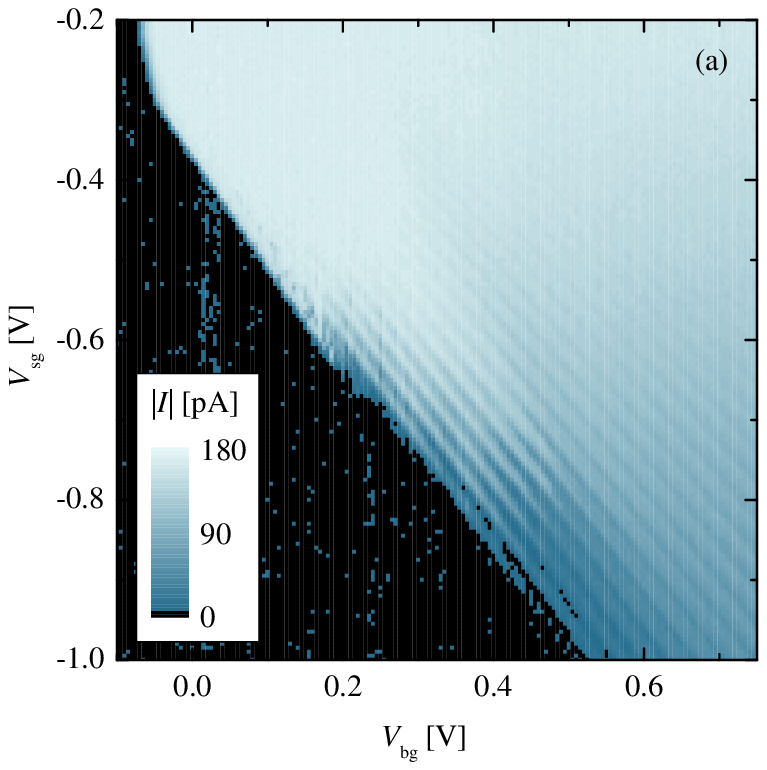}
 \includegraphics[angle=0,width=0.45\textwidth]{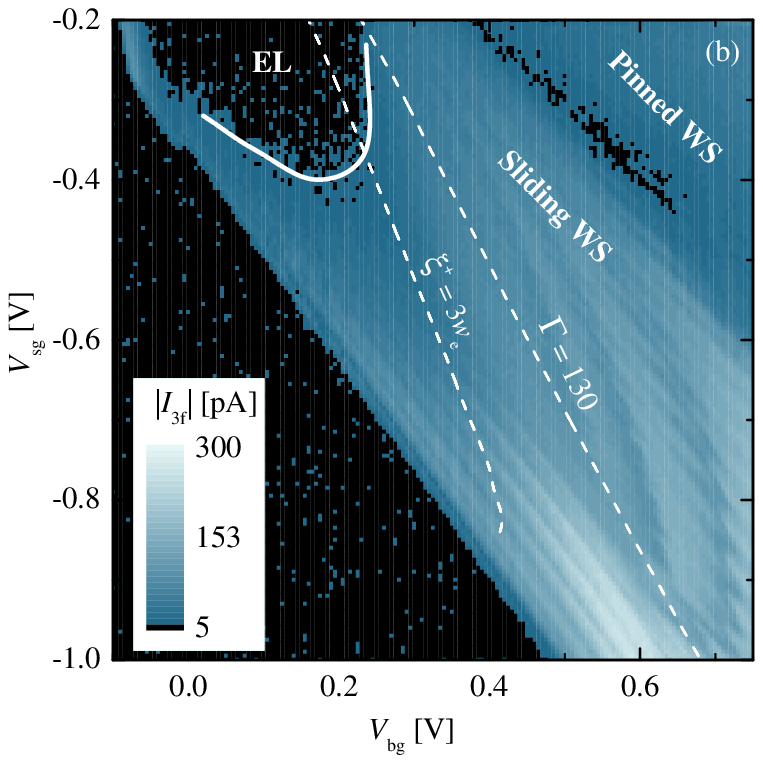}
 \caption{(a) Magnitude of $I$ against $V_{bg}$ and $V_{sg}$, at $T=0.6$ K. Here $V_{in}=3$ mV. (b) Magnitude of $I_{3f}$ measured under the same conditions, but with $V_{in}=10$ mV. The measurement noise floor was 5 pA. The solid line indicates the boundary between the electron liquid (EL) and WS phases. Regions corresponding to the EL phase and the sliding and dynamically pinned WS phases are labelled accordingly, as are lines representing the conditions $\Gamma=130$ and $\xi^+=3w_e$.  \label{Fig:4}}
\end{centering}
\end{figure}

From the current measurement shown in Fig. 2(a), for each point in the $V_{bg}$-$V_{sg}$ plane, it is not straightforward to determine whether the SSE system is in the electron liquid or Wigner solid phase. This is because there is no sharp contrast between high and low current regions. Although the current should drop when the electron system becomes dynamically pinned to the dimple lattice, the sinusoidal driving voltage can induce the decoupling the dimple lattice during each ac cycle, if the pinning effect is not sufficiently strong\cite{StickSlip}. In this case the measured current magnitude may not be greatly changed from that measured in the electron liquid phase. However, the nonlinear response in this transport regime gives rise to higher odd harmonics in the ac current signal\cite{ThirdHarmonic}. Therefore, in order to precisely determine the boundary between the liquid and solid phases, the third harmonic component of the SSE current, $I_{3f}$, was measured for varying $V_{bg}$ and $V_{sg}$. The result is shown in Fig. 2(b), for $V_{in}=10$ mV and $T=0.6$ K. 
In the upper left-hand corner of the plot, where $n_s$ is generally low, the ac response is linear signifying the electron liquid regime. In the Wigner solid regime, distinct regions in which $I_{3f}$ is high and low are evident. These correspond to the sliding and dynamically pinned transport regimes, respectively. In the sliding regime, the decoupling of the electron system from the dimple lattice during each ac current cycle leads to large $I_{3f}$. The decoupling occurs more readily close to the conduction threshold, where $n_s$ is low. In the dynamically pinned regime, in which $n_s$ is higher, the driving force cannot induce the decoupling and BC scattering limits the electron velocity during each ac cycle. This also results in a nonlinear SSE response, but with a reduced current magnitude and so smaller $I_{3f}$.  

In 2D, the Wigner solid melts when the value of $\Gamma$, which is defined as the ratio of the electron Coulomb energy $U_C=e^2\sqrt{\pi n_s}/4\pi\varepsilon_0$ to kinetic energy $k_B T$, falls below a critical value of $\Gamma^{2\text{D}}\approx130$. Here $e$, $\varepsilon_0$ and $k_B$ are the elementary charge, vacuum permittivity and the Boltzmann constant, respectively. The melting is a Kosterlitz-Thouless (KT)-type transition, occurring due to the unbinding of lattice dislocation pairs\cite{KosterlitzThouless1973,HalperinNelson1978,Young1979,Morf1979}. In a small temperature range above the melting temperature there exists a `hexatic' phase, in which the system exhibits nearest-neighbour bond-orientational order. This order is finally destroyed at a higher temperature by the emergence of a second type of defect, lattice disclinations, and the system enters the isotropic liquid phase. In this work we define melting as the loss of long-range positional order due to the appearance of free dislocations because, for electrons on helium, the dimple lattice formation and associated transport properties depend on this positional ordering. The distance over which positional ordering can be expected is described by the correlation length $\xi^+=ae^{b/t^{\nu}}$ where $a\approx n_s^{-0.5}$ is the dislocation core size, $b=1.8$ is the ratio of the core energy to the thermal energy, $\nu=0.37$ is a constant and $t=\frac{\Gamma^{2D}}{\Gamma}-1$. The correlation length is finite in the electron liquid phase and diverges as $\Gamma$ approaches $\Gamma^{2\text{D}}$. However, for quasi-1D systems, an ordered state might be expected when $\xi^+$ exceeds the system width $w_e$, which can occur for $\Gamma<\Gamma^{2\text{D}}$. Recently, the melting of a quasi-1D SSE system was found to be described by the empirical relation $\xi^+=3w_e$\cite{Ikegami2015}. We note that a strictly quantitative comparison between experiment and theory is difficult in this case because the values of $b$ and $\nu$ are valid only when $t\leq0.07$.   

 \begin{figure} 
\begin{centering}
   \includegraphics[angle=0,width=0.45\textwidth]{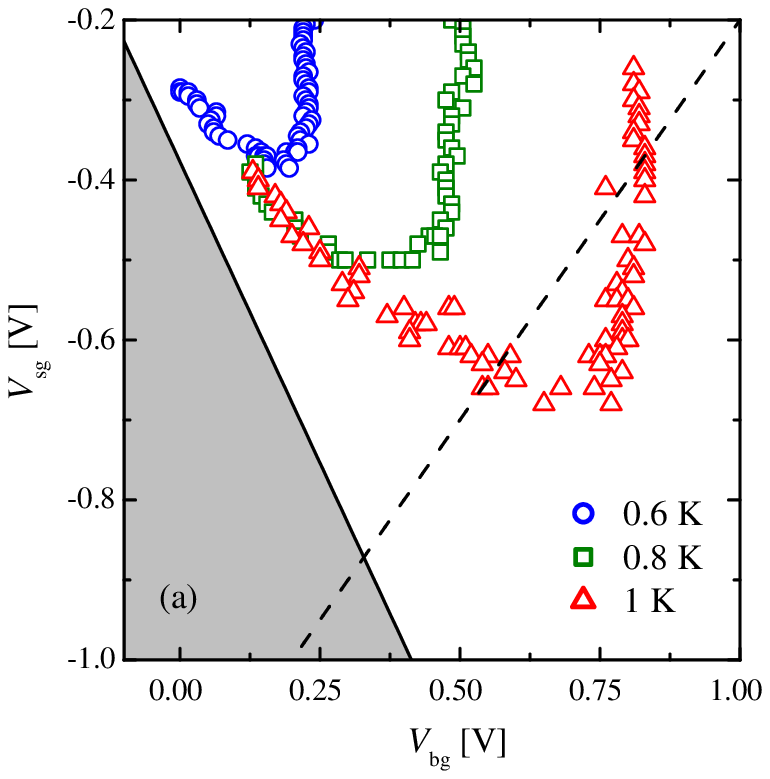}
  \includegraphics[angle=0,width=0.45\textwidth]{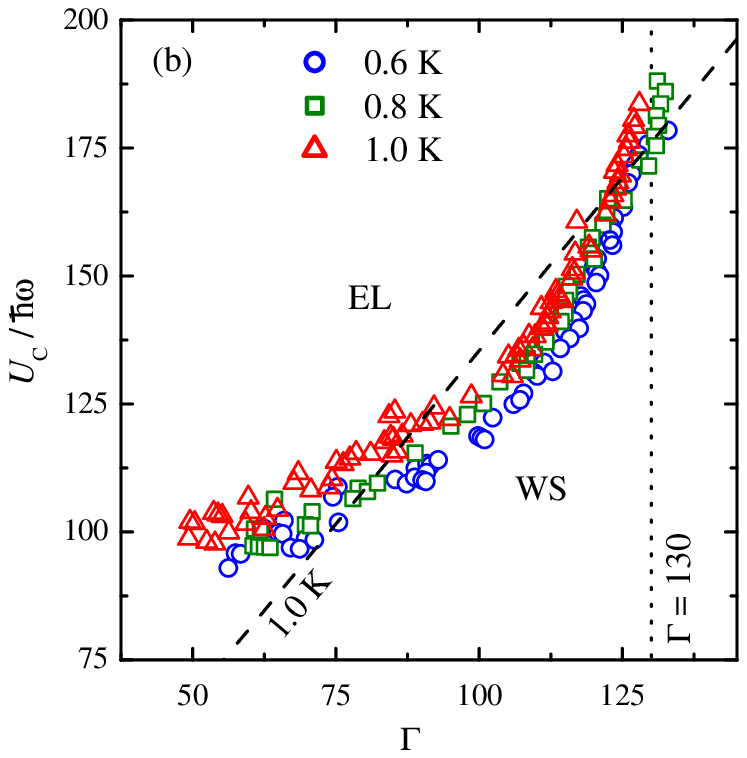}
 \caption{(a) Phase boundaries determined by measuring $I_{3f}$, at $T=0.6$, 0.8 and 1.0 K. The solid line represents the threshold for conductance through the central microchannel. Along the dashed line $\Delta V$ is constant. (b) $U_C/\hbar\omega$ against $\Gamma$ for the data shown in (a). The dotted line represents the 2D melting criterion. The dashed line is equivalent to the dashed line in (a), for $T=1$ K.\label{Fig:4}}
\end{centering}
\end{figure}

Lines representing the 2D and quasi-1D melting criteria are shown in Fig. 2(b). (In our calculations the screening of the Coulomb interaction by the underlying electrode, which modifies $U_C$ by a few per cent, is taken into account\cite{PhysRevB.82.201104, *Ikegami2015}.) When $V_{sg}$ is close to $V_{gu}$ the boundary between the electron liquid and solid regimes is close to the 2D melting criterion, confirming the KT-type melting. However, for more negative $V_{sg}$, the electron system is solid within (to the left of) both the $\Gamma=130$ and $\xi^+=3w_e$ boundaries. We conclude that electron ordering in quasi-1D generally occurs for values of $\Gamma$ below 130 and, in our sample, is found to be more strongly enhanced than in previous experiments where the phase boundary followed the condition $\xi^+=3w_e$\cite{Ikegami2015}.

Phase boundaries for $T=0.6$, 0.8 and 1.0 K are shown in Fig. 3(a). The boundaries were determined by finding the values of $V_{bg}$ and $V_{sg}$ for which $I_{3f}$ exceeds the measurement noise floor, after adjacent-point averaging over a 50 mV window. The liquid region expands as $T$ increases, as expected. To elucidate the relationship between $U_C$, $\omega$ and $T$ at the phase boundaries, we show in Fig. 3(b) the dimensionless ratio $U_C/\hbar\omega$ against $\Gamma$. (It is convenient to express the confinement in units of energy although no quantum mechanical effects are considered here). The data points fall close to a single curve, suggesting that $U_C$, $\omega$ and $T$ are interrelated at the Wigner solid melting point. When $U_C/\hbar\omega$ is large the melting occurs close to the 2D criterion $\Gamma=130$. As $U_C/\hbar\omega$ decreases, due to increasing confinement strength or decreasing electron density, the Wigner solid melts at values of $\Gamma$ much lower than 130. This observation is in agreement with numerical simulations, which have shown that strong electrostatic confinement restricts lateral particle motion and thereby suppresses the melting of the quasi-1D Wigner solid\cite{Peeters1DCrystal,PackingAndMelting}. The dependence of melting on confinement strength is qualitatively different from finite-size effects considered in other studies\cite{PhysRevB.82.201104, *Ikegami2015}. We conclude that the KT-type melting of the Wigner solid can be significantly modified by strong lateral confinement, as it promotes the positional order of the electron system. To our knowledge, this is the first evidence of a scaled relationship between interaction energy, confinement strength and temperature that describes the melting of a quasi-1D system.  

The interplay between $U_C$, $\omega$ and $T$ at the Wigner solid melting point results in a curvature of the phase boundaries shown in Figs. 3(a) and 3(b). Lines drawn on these plots for constant confinement voltage $\Delta V=V_{bg}-V_{sg}$, and therefore constant $\omega$, can intersect the phase boundary twice, for a given temperature. Moving along these lines, by increasing $n_s$ under the constant electrostatic confinement, therefore results in a reentrant solid-liquid-solid transition. We are aware of no previous demonstration of such behaviour, but consider it likely to be observed in other strongly correlated quasi-1D systems subjected to parabolic-like confinement. 

\begin{figure} 
\begin{centering}
 \includegraphics[angle=0,width=0.45\textwidth]{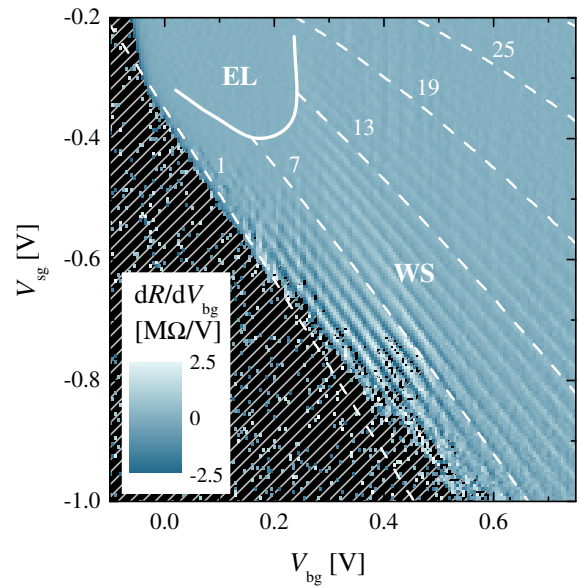}
 \caption{d$R$/d$V_{bg}$ against $V_{bg}$ and $V_{sg}$ at $T=0.6$ K and $V_{in}=3$ mV. In the hatched region $I$ becomes too small to measure. The dashed lines represent constant values of $N_y$ as given by the FEM analysis. The solid line represents the liquid-solid boundary determined by the measurement of $I_{3f}$, as also shown in Fig. 2(b).\label{Fig:2}}
\end{centering}
\end{figure}

In Fig. 2, both $I$ and $I_{3f}$ exhibit fringe-like features that lie close to parallel with the conductance threshold. Previous studies have shown that such oscillatory transport behaviour arises due to the modulation of the electron lattice structural order with changing $N_y$\cite{PhysRevLett.109.236802, *rees2013reentrant}. Close to the structural transitions between adjacent $N_y$, lattice defects and fluctuations between lattice configurations of similar energy should occur\cite{Peeters1DCrystal}. The reduced positional order weakens the Bragg ripplon scattering, and so the effective electron mass and the SSE resistivity. A plot of d$R$/d$V_{bg}$ for varying $V_{bg}$ and $V_{sg}$ is shown in Fig. 4. Oscillatory resistance features follow arc-like paths in the $V_{bg}$-$V_{gt}$ plane. No resistance oscillations are recorded in the electron liquid region. Several lines corresponding to constant $N_y$ values, as calculated by the FEM for $\phi_e=-0.151$ V, are also shown. These lines closely follow the $R$ maxima, confirming that the resistance oscillations are related to the changing $N_y$. However, this effect was previously only observed close to the melting point of the Wigner solid and for small $N_y$\cite{PhysRevLett.109.236802, *rees2013reentrant}. Here we demonstrate that structural transitions strongly influence the electron positional order up to large $N_y$ and for $\Gamma \gg 130$.

The solid-liquid-solid transition that occurs when $n_s$ increases for certain values of $\Delta V$ results in the loss of the resistance oscillations for intermediate values of $N_y$. In Fig. 5 we show both $R$ and $I_{3f}$ recorded along the line in the $V_{bg}$-$V_{sg}$ plane for which $\Delta V=0.55$ V, at $T=0.6$ K. In both measurements, signatures of electron ordering are exhibited for small and large $N_y$ but not for intermediate values $5 \lesssim N_y \lesssim 13$, confirming the reentrant behaviour. The close agreement between the two independent measurements, performed for different $V_{in}$, confirms that the phase boundary is accurately determined in our transport measurements. 

\begin{figure} 
\begin{centering}
 \includegraphics[angle=0,width=0.45\textwidth]{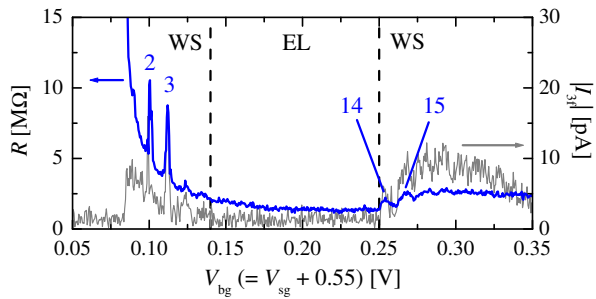}
 \caption{$R$ and the magnitude of $I_{3f}$ recorded along the line $\Delta V=V_{bg}-V_{sg} = 0.55$ V, at $T=0.6$ K. The electron liquid and Wigner solid regimes, and resistance peaks corresponding to several $N_y$, are labelled. \label{Fig:2}}
\end{centering}
\end{figure}

\subsection{C. Monte Carlo Simulations}
To better understand the nature of the structural transitions that occur as $N_y$ increases, we performed Monte Carlo simulations of electrons interacting via a screened Coulomb potential under parabolic confinement. Techniques similar to those described in Ref.\cite{Peeters1DCrystal} were used, but extended to larger $N_y$. We consider $N$ classical particles in 2 dimensional space interacting with each other through a Yukawa potential. The particles are confined in the $y$ direction by a parabolic potential, and there is no confinement along the $x$ axis. Periodic boundary conditions are inserted in the $x$ direction to simulate an infinite length in $x$. We solved the Langevin equation with a friction force proportional to the velocity and a temperature dependent random force. The ground state structures at zero temperature were searched for by the annealing method. The number of particles required to achieve the smallest energy per particle were found by repeating the annealing simulation for various $N$, with the box length adjusted to keep the linear density constant. In this way the proper number of particles to obtain the lowest-energy structures could be determined. We used the scaled Hamiltonian used in Ref.\cite{Peeters1DCrystal}, with $\kappa =7.25$. This corresponds to a screening length $\lambda=1$ $\SI{}{\micro\meter}$ for an electron system under parabolic confinement $\hbar\omega/k_B=0.6$ K. These values are comparable with the experimental conditions.   

\begin{figure} 
\begin{centering}
\includegraphics[angle=0,width=0.5\textwidth]{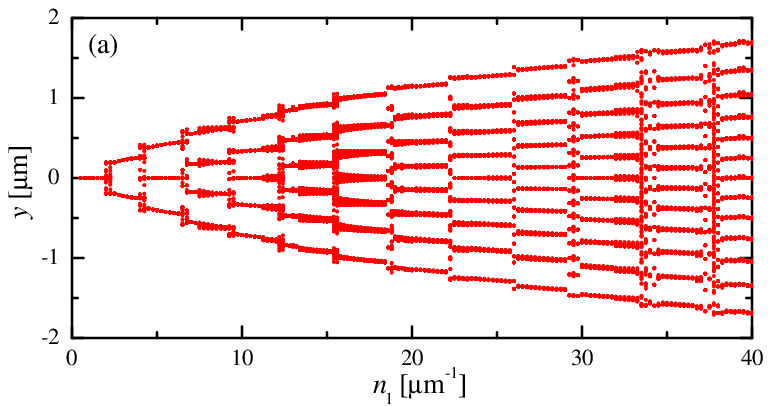}
\includegraphics[angle=0,width=0.18975\textwidth]{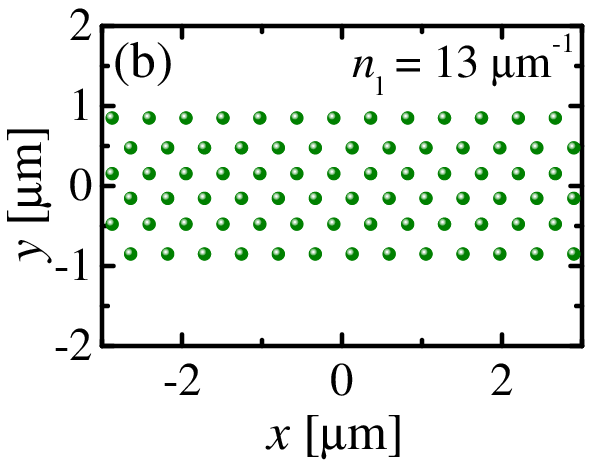}
\includegraphics[angle=0,width=0.253\textwidth]{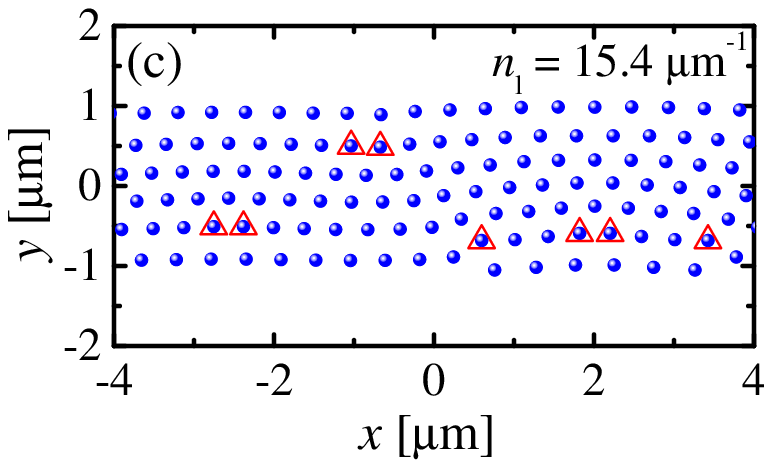}
\caption{(a) $y_i$ against $n_l$ as calculated by Monte Carlo simulations. (b) Electron lattice configuration for $n_l=13$ $\SI{}{\micro\meter}^{-1}$. The simulation cell contains 84 particles. (c) Electron lattice configuration for $n_l=23.5$ $\SI{}{\micro\meter}^{-1}$. The simulation cell contains 141 particles. Electrons exhibiting 5 nearest neighbours are marked by triangles.\label{Fig:3}}
\end{centering}
\end{figure} 

Lateral particle positions $y_i$ against linear electron density $n_l$ are shown in Fig. 6(a). Generally, the electrons are arranged in rows and $N_y$ increases sequentially. An example of a highly ordered row structure, found for $n_l=13$ $\SI{}{\micro\meter}^{-1}$, is shown in Fig. 6(b). However, close to values of $n_l$ at which $N_y$ changes, an increased scatter in $y_i$  reflects reduced positional order. An example is shown in Fig. 6(c) for $n_l=15.4$ $\SI{}{\micro\meter}^{-1}$; the electron lattice becomes distorted and domains containing 6 or 7 rows appear, along with structural defects that break the six-fold symmetry of the electron lattice. This behaviour is in agreement with other similar studies made for small values of $N_y$\cite{PackingAndMelting}. The increased disorder observed here occurs at each transition and up to the largest $N_y$ values, supporting our explanation that the resistance oscillations observed in the experiment reflect changes in $N_y$ as $V_{bg}$ or $V_{sg}$ is varied.

\section{III. CONCLUSION}
We have used a multigated microchannel device to map the first structural and phase diagrams for a quasi-1D electron system on a liquid He substrate.  The KT-type Wigner solid melting is strongly modified by the lateral electrostatic confinement. We have demonstrated for the first time that the melting of the quasi-1D Wigner solid is determined by a scaled relationship between Coulomb energy, temperature and the confinement strength. In addition we have shown that the positional order of electrons in the quasi-1D Wigner solid depends strongly on the commensurability with the confinement geometry, even when the number of electron rows is large. This observation was confirmed using Monte Carlo simulations. Because electrons on helium are a model system, our results are relevant to a wide variety of micro- and macroscopic many body systems.     

\section{ACKNOWLEDGEMENTS}
We thank A. D. Chepelianskii and Yu. Lysogorskiy for helpful discussions. This work was supported by JSPS KAKENHI Grant No. JP24000007, and by the Taiwan Ministry of Science and Technology (MOST) through Grant Nos. MOST 103-2112-M-009-001, MOST 103-2112-M-009-017 and MOST 102-2112-M-009-014-MY2, and by the MOE ATU Program. This work was performed according to the Russian Government Program of Competitive Growth of Kazan Federal University.


%

\end{document}